# Nonlocal radiative coupling in non monotonic stellar winds

A. Feldmeier and R. Nikutta

Astrophysik, Institut für Physik, Universität Potsdam, Am Neuen Palais 10, 14469 Potsdam, Germany  
e-mail: `afeld, robert@astro.physik.uni-potsdam.de`

**Abstract.** There is strong observational evidence of shocks and clumping in radiation-driven stellar winds from hot, luminous stars. The resulting non monotonic velocity law allows for radiative coupling between distant locations, which is so far not accounted for in hydrodynamic wind simulations. In the present paper, we determine the Sobolev source function and radiative line force in the presence of radiative coupling in spherically symmetric flows, extending the geometry-free formalism of Rybicki and Hummer (1978) to the case of three-point coupling, which can result from, e.g., corotating interaction regions, wind shocks, or mass overloading. For a simple model of an overloaded wind, we find that, surprisingly, the flow decelerates at all radii above a certain height when nonlocal radiative coupling is accounted for. We discuss whether radiation-driven winds might in general not be able to re-accelerate after a non monotonicity has occurred in the velocity law.

## 1. Introduction

In a spherically symmetric, *accelerating* stellar wind, each point recedes from any other one. A photospheric photon can therefore be scattered in a given spectral line of infinitesimal width only at one specific radius where the photon frequency matches the Doppler-shifted line transition frequency. We refer to this situation as one-point coupling. For strong acceleration, the scattering takes place in a narrow spherical layer, the so-called Sobolev zone. This zone has a width of a few Sobolev lengths $L = v_{\rm th}/(dv/dr) \ll H$, with radius $r$, wind speed $v$, thermal speed $v_{\rm th}$, and scale height $H = -n/(dn/dr)$ with respect to the density $n$ of the line-forming species. Pressure broadening is neglected. In the Sobolev approximation, the density $n$ and velocity gradient $dv/dr$ are assumed to be constant within the Sobolev zone. Equivalently, in the limit $L \to 0$, the Sobolev zone becomes a mathematical surface.

In a spherically symmetric, expanding, *decelerating* medium, a photon can be scattered repeatedly at different radii within a single line transition. For each scattering location $r$, these resonance locations $r'$ form a *resonance surface* in the limit $L \to 0$ (Rybicki & Hummer 1978; RH in the following). For a velocity law $v \sim 1/r$, Fig. 1 shows a cut through the resonance surface. Each straight line through $r$ cuts the resonance surface in one other point at most. This is a consequence of the surface being convex. We refer to this as two-point coupling ($r$ and $r'$).

Rybicki and Hummer derive an equation for the Sobolev source function $S$ in the presence of multiple scattering. For each radius $r$ and every photon direction $\mathbf{n}$, the coupling lo-cations $r', r'', \ldots$ are determined first. Then the integral equation for $S$, which results when adding the contributions from all coupling directions, is solved by lambda iteration. The latter converges quickly if the number of disjunct resonances in each direction is small. This formalism was applied by Puls et al. (1993) for line-profile synthesis in non monotonic wind velocity laws resulting from the line deshadowing instability.

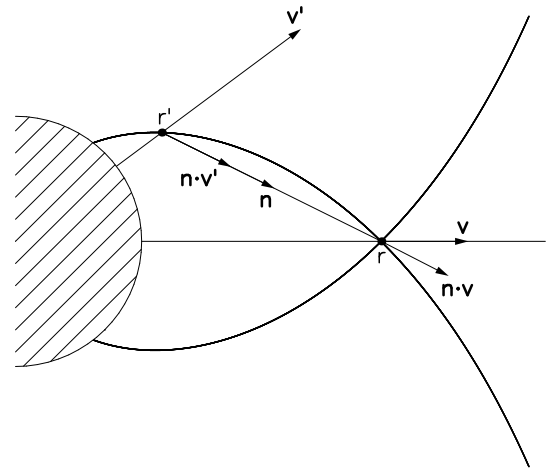

**Fig. 1.** A cut through the resonance surface for a decelerating velocity law $v \sim r^{-1}$. The resonance surface is obtained by rotating the figure about the symmetry axis. The projected velocity $\mathbf{n} \cdot \mathbf{v}'$ along direction $\mathbf{n}$ from any point $r'$ on the resonance surface towards $r$ equals the projected velocity $\mathbf{n} \cdot \mathbf{v}$ at $r$.

For two-point coupling in a monotonically decelerating, spherically symmetric flow, RH derive a "geometry-free" equation for $S$ that shows no reference to the shape of the resonance surface. This is achieved by a variable substitution $\mu \to r'$, with $\mu$ the cosine of the angle between the radial direction **r** and the coupling direction **n**.

There is a number of possible reasons why stellar wind velocity laws could become non monotonic and show multiple photon scattering in line transitions:

(1) *DACs*. So-called discrete absorption components migrate through the absorption trough of unsaturated P Cygni line profiles from hot, massive stars (Kaper & Fullerton 1998, and references therein). They are expected to originate in clumps in the rotating stellar wind, which generate upstream propagating radiative-acoustic waves (Cranmer & Owocki 1996). The wind decelerates between the wave head (a kink in the velocity law) and the clump, and accelerates elsewhere.

(2) *Backfalling clumps*. Clumps of gas falling back to the photosphere were proposed as the origin of redshifted absorption features in line profiles from $\tau$ Sco (Lamers & Rogerson 1978), as well as of its hard X-ray emission (Howk et al. 2000). The wind velocity law is non monotonic in the presence of backfalling clumps.

(3) *Line-driven instability*. The instability of radiative driving by scattering in spectral lines is thought to be responsible for the formation of strong wind shocks (Lucy & Solomon 1970; Owocki et al. 1988), making the velocity law highly non monotonic.

(4) *Ion runaway*. Recent simulations of thin winds show that radiation-driven metal ions do not decouple from the bulk H and He plasma as suggested by Castor et al. (1976), but that H, He, and metal ions switch together as a single fluid to a coasting solution (Krtička & Kubát 2000). Some of the velocity laws obtained in this way are again non monotonic.

(5) *Wind choking*. Multiple critical points (Holzer 1977) may occur in stellar winds for different physical reasons, e.g. due to a complicated run of the radiative flux with radius resulting from ionization changes. This may cause choking and mass overloading of the flow, leading to a non monotonic velocity law.

In non monotonic winds that accelerate everywhere except within a finite radial interval, up to three resonance locations appear for radially propagating photons and, likewise, in lateral directions. The present paper generalizes the geometry-free approach of RH to this case of three-point coupling. We emphasize that spherical symmetry is assumed throughout. An alternative method to solve the one-dimensional spherical radiative transfer for non monotonic flows was published by Baron & Hauschildt (2004), using accelerated lambda iteration in the comoving frame. We compare both approaches in Section 8.

## 2. One-point coupling

The Sobolev approximation for steeply accelerating or decelerating flows is developed in detail in RH, Rybicki (1970), and Lucy (1971). Here we summarize the main results without derivation. Complete redistribution is assumed for a two-level atom. Let $\phi(x)$ be the Doppler line profile function, with $x$ the distance from line center in Doppler units measured in a frame comoving with the atom, and $\mu = \cos\theta$ be the direction cosine as introduced above. The stellar photospheric intensity $I^*_\nu$ shall be angle-independent and constant over the line profile. We assume that the mass absorption coefficient $\kappa$ (units cm$^2$/g) and the thermal speed $v_{\rm th}$ are independent of $r$. The Sobolev optical depth $t$ is given by

$$t(r,\mu,x) = \tau(r,\mu)\Phi(x), \qquad (1)$$

where

$$\Phi(x) = \int_x^\infty dx'\,\phi(x'), \qquad (2)$$

$$\tau(r,\mu) = \kappa v_{\rm th}\varrho(r)/|Q(r,\mu)|, \qquad (3)$$

with velocity gradient in direction $a\cos\mu$,

$$Q(r,\mu) = \mu^2 \frac{dv}{dr} + (1-\mu^2)\frac{v}{r}, \qquad (4)$$

and density $\varrho(r)$. The formal solution for the line intensity is, with source function $S$

$$I_\nu(r,\mu,x) = I^*_\nu D(\mu)e^{-\tau(r,\mu)\Phi(x)} + S_\nu(r)\left(1 - e^{-\tau(r,\mu)\Phi(x)}\right), \qquad (5)$$

where $D = 1$, if the ray hits the stellar core, and 0 otherwise. Performing the double integral $\int d\mu \int dx\,\phi(x)$ over angle and frequency gives the mean line intensity,

$$\bar{J}_\nu(r) = I^*_\nu \beta_{\rm c}(r) + S_\nu(r)[1 - \beta(r)], \qquad (6)$$

with escape probability

$$\beta(r) = \frac{1}{2}\int_{-1}^{1} d\mu \frac{1 - e^{-\tau(r,\mu)}}{\tau(r,\mu)}, \qquad (7)$$

and, correspondingly for $\beta_{\rm c}$, with an extra factor $D(\mu)$ under the integral. For pure scattering, $S_\nu = \bar{J}_\nu$, and

$$S_\nu(r) = \frac{\beta_{\rm c}(r)}{\beta(r)} I^*_\nu. \qquad (8)$$

## 3. Two-point coupling

Rybicki and Hummer generalize the Sobolev method to the case that up to $n$ disjunct radiative couplings occur in any photon propagation direction. Most relevant to the present paper is the geometry-free approach of RH for $n = 2$.

To account for photons originating either from the stellar photosphere at radius $R_*$ or from some location $r'$ on the resonance surface, RH replace, in Eq. (5),

$$I^*_\nu D(\mu) \to I^*_\nu D(\mu) e^{-\tau(r',\mu')} + S_\nu(r')\left(1 - e^{-\tau(r',\mu')}\right). \qquad (9)$$

Note that no $\Phi(x)$ appears in this expression, since only photons emerging from the resonance layer towards $r$ matter. Hence, $x = -\infty$ and $\Phi = 1$. Let $\mu = \mathbf{r}\cdot\mathbf{n}$ and $\mu' = \mathbf{r}'\cdot\mathbf{n}$, with photon propagation direction **n**. Inserting Eq. (9) into Eq. (5), and performing $\int d\mu \int dx\,\phi(x)$, one finds

$$\bar{J}_\nu(r) = I^*_\nu \tilde{\beta}_{\rm c}(r) + \frac{1}{2}\int_{-1}^{1} d\mu \left(1 - e^{-\tau'}\right)\frac{1 - e^{-\tau}}{\tau} S_\nu(r') \\ + S_\nu(r)[1 - \beta(r)], \qquad (10)$$

where $\tau = \tau(r,\mu), \tau' = \tau(r',\mu')$, and $\tilde{\beta}$ differs from $\beta$ in Eq. (7) by an extra factor $e^{-\tau'}$ under the integral. For pure scattering,

$$S_\nu(r) = \frac{\tilde{\beta}_c(r)}{\beta(r)} I_\nu^* + \frac{1}{2\beta(r)} \int_{-1}^{1} d\mu \left(1 - e^{-\tau'}\right) \frac{1 - e^{-\tau}}{\tau} S_\nu(r'). \quad (11)$$

Here, $S_\nu(r')$ cannot be taken outside the integral, since the source function varies along the resonance surface, $r' = r'(\mu)$.

For given **n**, the resonance condition between radii $r$ and $r'$ is $\mu v = \mu' v'$, where $v = v(r)$ and $v' = v(r')$. The latter is not to be confused with the spatial derivative $dv/dr$. If $p$ is the normal from the coordinate origin onto the photon ray, $p = r\sqrt{1-\mu^2} = r'\sqrt{1-\mu'^2} = p'$ holds. Together with the resonance condition, this allows us to solve for $\mu, \mu'$ in terms of the resonance locations $r, r'$,

$$\mu^2 = \frac{(r'^2 - r^2)v'^2}{r'^2 v^2 - r^2 v'^2}, \quad (12)$$

and similarly for $\mu'$, with $r, v$ replaced by $r', v'$, and vice versa. With Eqs. (3) and (4),

$$\tau = \kappa v_{\text{th}} \varrho(r) \frac{r(r'^2 v^2 - r^2 v'^2)}{r(r'^2 - r^2)v'^2 dv/dr + vr'^2(v^2 - v'^2)}, \quad (13)$$

and similarly for $\tau'$. In the following, $\tau, \tau'$ are understood either as a function of $r$ and $\mu$ or as a function of $r$ and $r'$, depending on whether they appear in integrals over $\mu$ or $r'$, respectively. From Eq. (12), $d\mu/dr$ can be calculated and then used for a variable substitution $\mu \to r'$ in Eq. (11), giving (see RH)

$$S_\nu(r) = \frac{\tilde{\beta}_c(r)}{\beta(r)} I_\nu^* + \frac{\kappa v_{\text{th}}}{2\beta(r)} \int_{R_*}^{\infty} dr' r'^2 \varrho(r') S_\nu(r') K(r, r'), \quad (14)$$

with

$$K(r, r') = \frac{1 - e^{-\tau}}{\tau} \frac{1 - e^{-\tau'}}{\tau'} \Upsilon_1 \quad (15)$$

and

$$\Upsilon_1 = \frac{1}{\sqrt{(r^2 - r'^2)(r^2 v'^2 - r'^2 v^2)}}. \quad (16)$$

Note that $K$ is symmetric in $r$ and $r'$. The integral equation (14) can be solved using lambda iteration. Accelerated lambda iteration is not required, as $S$ converges quickly. Equations (14) and (15) show no reference to the shape $\mu(r, r')$ of the resonance surface. A Taylor series expansion to first order in powers of $r' - r$ shows that $K(r, r) = 0$. This relation is important in numerical implementations of the RH method.

## 4. Three-point coupling

Monotonically falling wind velocity laws have rather limited physical relevance. We now consider velocity laws that accelerate everywhere, except for a single, finite region of deceleration. Figure 2 shows the corresponding topology of resonance surfaces, where we have chosen $v \sim r$ for $r < 2$ and $r > 3$, and $v \sim 1/r$ in between. Kinks, i.e. discontinuities in $dv/dr$, occur at $r_1 = 2$ and $r_2 = 3$. Three cases can be distinguished according to whether $r$ lies in the acceleration region at small radii (case I), in the deceleration region (case II), or in the acceleration region at large radii (case III).

The resonance surface has sharp cusps at the location of the kinks. For case II, the interval $[r_1, r_2]$ corresponds to a part of the RH resonance surface in Fig. 1, and is termed *lateral coupling region* in the following. The regions $r < r_1$ and $r > r_2$ that close the radiative coupling surface will be referred to as *radial coupling regions* or *resonance caps*. The points $r_-$ and $r_+$ on the resonance surface are for purely radial coupling, where $\mu = \mu' = 1$. Note that $r_-$ and $r_+$ depend on $r$. For cases I and III, $r$ is disjunct to the resonance surface. Let $R_- = \min_r\{r_-\}, R_+ = \max_r\{r_+\}$ be the first and last points in the wind with multiple radiative coupling. In addition, inflection points in the lateral coupling region may cause five-point coupling. This affects only locations $r$ very close to kinks and thus just a small portion of the *projected* resonance surface as seen from $r$. We therefore neglect five-point coupling throughout.

A central aspect of the present paper is that Eq. (14) derived by RH for the case of two-point coupling also holds for three-point coupling if the integral boundaries are chosen appropriately. We consider case II first. Equation (14) was derived under the assumption that along each photon *ray* through $r$, there is one disjunct resonance point. The equation remains valid for three-point coupling, i.e. if for each *direction* from $r$ there is one disjunct resonance. The function $\mu(r')$ is monotonic over the full resonance surface if no inflection points occur and then substitution of $r'$ for $\mu$ is possible. One easily sees that inflection points on the lateral coupling region pose no difficulties in the substitution of $r'$ for $\mu$. Therefore, Eq. (14) holds for case II of three-point coupling when the integration boundaries $R_*$ and $\infty$ are replaced by $r_-$ and $r_+$, respectively. Since $r_-$ and $r_+$ are the first and last *radial* coupling location, they can be determined from $v(r)$ without reference to the shape of the resonance surface. Cases I and III can be treated similarly, although two resonances exist along a given direction $\cos^{-1} \mu$ from $r$. To sample the resonance surface, two separate $\mu$ integrals have to be performed over the same $\mu$ range. As is clear from Fig. 2, they can be replaced by a single $r'$ integral from $r_-$ to $r_+$. The geometry-free generalization of the RH formalism to three-point coupling is therefore

$$S_\nu(r) = \frac{\tilde{\beta}_c(r)}{\beta(r)} I_\nu^* + \frac{\kappa v_{\text{th}}}{2\beta(r)} \int_{r_-}^{r_+} dr' r'^2 \varrho(r') S_\nu(r') K(r, r'), \quad (17)$$

with

$$K(r, r') = \frac{1 - e^{-\tau}}{\tau} \frac{1 - e^{-\tau'}}{\tau'} e^{-\tau''} \Upsilon_1. \quad (18)$$

The tilde in $\tilde{\beta}_c$ refers now to extinction of stellar photons at 0, 1, or 2 intervening resonances, depending on $\mu$ and whether case I, II, or III is considered. In the kernel (18), $\exp(-\tau'')$ accounts for extinction at $r''$ of photons originating from $r'$. The kernel is still symmetric, since $\tau''(r, r') = \tau''(r', r)$. All extinctions are easily determined from the resonance conditions by means of numerical root search. Symmetry relations allow the radial search domain to be constrained.

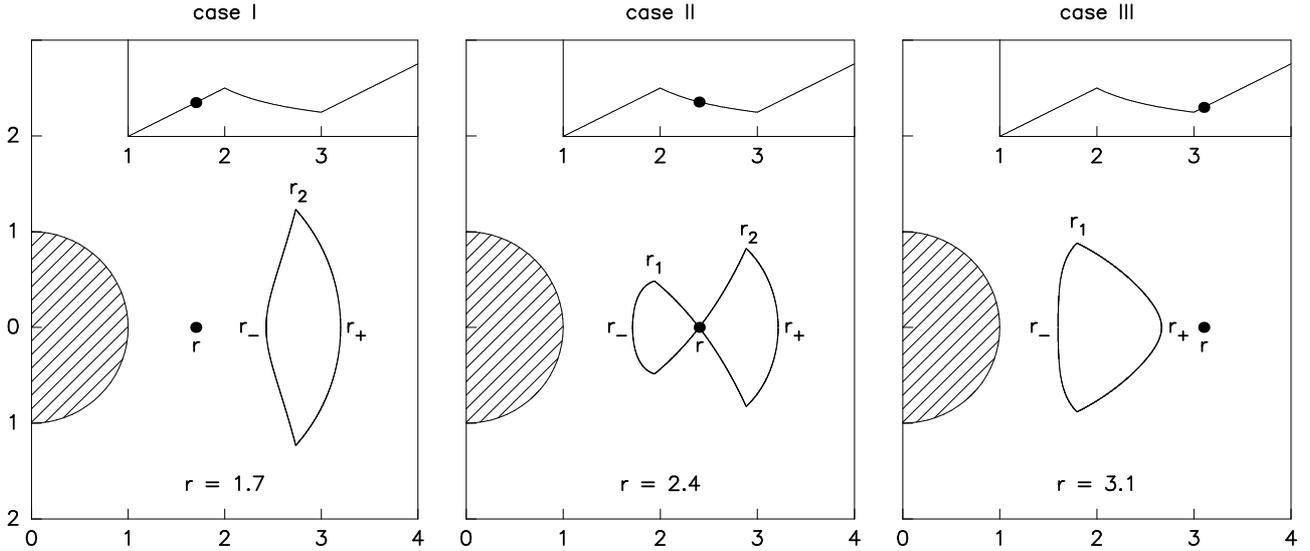

**Fig. 2.** Cut through resonance surfaces of a non monotonic test velocity law (shown in the subpanels) defining cases I, II and III.

Once $S_\nu$ is known, intensity $\bar I_\nu$ is obtained from a formal solution of the transfer equation. The angle integral in the calculation of the flux $\bar H_\nu$ is again transformed to a spatial integral. The generalization of the relevant Eq. (71) of RH to three-point coupling is,

$$\bar H_\nu(r) = \tilde\gamma_c(r) I_\nu^* + \frac{\kappa v_{\rm th}}{2}\int_{r_-}^{r_+} dr'\, r'^2 \varrho(r') S_\nu(r') L(r,r'), \quad (19)$$

with

$$\tilde\gamma_c(r) = \frac{1}{2}\int_{-1}^{1} d\mu\,\mu D(\mu)\, e^{-\tau'}\,\frac{1-e^{-\tau}}{\tau}, \quad (20)$$

and kernel

$$L(r,r') = \frac{1-e^{-\tau}}{\tau}\,\frac{1-e^{-\tau'}}{\tau'}\, e^{-\tau''} \Upsilon_2, \quad (21)$$

introducing

$$\Upsilon_2 = \frac{v'}{r^2 v'^2 - r'^2 v^2}. \quad (22)$$

## 5. The line force

The acceleration of gas due to radiation scattering in a single spectral line (subscript "l") of frequency $\nu$ and mass absorption coefficient $\kappa$ is, for radial flow and assuming pure Doppler broadening (see Mihalas 1978),

$$g_{\rm l}(r) = \frac{4\pi\kappa v_{\rm th}}{c^2}\,\nu \bar H_\nu(r), \quad (23)$$

with $\bar H_\nu$ from Eq. (19). To calculate the acceleration from an ensemble of non-overlapping lines (subscript "L"), we use the line distribution function (Castor et al. 1975),

$$N(\nu,\kappa) = \frac{1}{\nu\kappa_0}\left(\frac{\kappa_0}{\kappa}\right)^{2-\alpha}. \quad (24)$$

Here, $0 < \alpha < 1$, and $\kappa_0$ refers to the strongest line in the flow. Integrating Eq. (23) over $\nu$ and $\kappa$ using Eqs. (19) and (24),

$$g_{\rm L}(r) = \frac{2\pi v_{\rm th}}{c^2}\kappa_0^{1-\alpha}\Bigg[\int_{-1}^{1} d\mu\,\mu D(\mu)\frac{I_*}{\eta}\int_0^\infty d\kappa\,\kappa^{\alpha-2}\left(1-e^{-\kappa\eta}\right)e^{-\kappa\eta'}$$
$$+ v_{\rm th}\int_{r_-}^{r_+} dr'\, r'^2 \frac{\varrho'\Upsilon_2}{\eta\eta'}\int_0^\infty d\kappa\,\kappa^{\alpha-2} S(r')(1-e^{-\kappa\eta})(1-e^{-\kappa\eta'})e^{-\kappa\eta''}\Bigg], \quad (25)$$

where $I_* = \int_0^\infty d\nu I_\nu^*$, $S = \int d\nu S_\nu$, and $\eta(r,\mu) = \tau(r,\mu)/\kappa$. Furthermore, $\eta' = \eta(r',\mu')$ and $\tau_0(r,\mu) = \kappa_0\eta(r,\mu)$. We left $S$ under the $\kappa$ integral since the source function will be different for optically thin and thick lines. Using the mean value theorem, we now introduce an effective source function $S_{\rm e}$ independent of $\kappa$, which leaves the above integral unchanged (Owocki & Puls 1996).

A strong implicit assumption underlies Eq. (25). At radius $r$, a line of frequency $\nu$ scatters photons that were previously scattered at location $r'$ of the resonance surface by *all* lines lying within a few Doppler widths of $\nu$. Hence, independent integrals should be performed over $\kappa$ at $r$ and $\kappa'$ at $r'$. To simplify matters, we follow Castor et al. (1975) instead and assume a frequency separation $\geq v_\infty \nu/c$ between neighboring lines. In this case, Eq. (25) holds.

Since $\eta$ is independent of $\kappa$, the above $\kappa$ integrals can be performed analytically, using two formulae derived by an integration by parts,

$$\int_0^\infty d\kappa\,\kappa^{\alpha-2}\left(1-e^{-\kappa\eta}\right)e^{-\kappa\eta'} = \frac{\Gamma(\alpha)}{1-\alpha}\left[(\eta+\eta')^{1-\alpha} - \eta'^{1-\alpha}\right], \quad (26)$$

$$\int_0^\infty d\kappa\,\kappa^{\alpha-2}\left(1-e^{-\kappa\eta}\right)\left(1-e^{-\kappa\eta'}\right)e^{-\kappa\eta''} = \frac{\Gamma(\alpha)}{1-\alpha} \times$$
$$\times\left[(\eta+\eta'')^{1-\alpha} + (\eta'+\eta'')^{1-\alpha} - (\eta+\eta'+\eta'')^{1-\alpha} - \eta''^{1-\alpha}\right], \quad (27)$$

with the complete Euler Gamma function $\Gamma$. Introducing dimensionless quantities $r = r/R$, $v = v/V$, and $\rho = \varrho/\Pi$, where $R, V, \Pi$ are appropriate scales, one finds the following two equations

$$\begin{Bmatrix} \dfrac{g_L(r)}{g_*} \Xi - \tilde{\gamma}_{Lc}(r) \\ \dfrac{S_e(r)\tilde{\beta}_L(r)}{I_*} - \tilde{\beta}_{Lc}(r) \end{Bmatrix} = \dfrac{1}{2}\kappa_0 v_{th} \dfrac{R\Pi}{V} \int_{r_-}^{r_+} dr' r'^2 \rho' \begin{Bmatrix} \Psi_1 \\ \Psi_2 \end{Bmatrix} \dfrac{S_e(r')}{I_*} \times$$
$$\times \dfrac{1}{\tau_0 \tau_0'} \left\{ (\tau_0 + \tau_0'')^{1-\alpha} + (\tau_0' + \tau_0'')^{1-\alpha} - (\tau_0 + \tau_0' + \tau_0'')^{1-\alpha} - \tau_0''^{1-\alpha} \right\}.$$
(28)

The integral equation for $S_e$ was obtained by performing $\int d\nu\,\nu \int d\kappa\,\kappa\,N(\nu,\kappa)$ upon Eq. (17). The extra $\kappa$ is chosen to match the definition of $S_e$. In Eq. (28), $\Psi_{1,2} = \Upsilon_{1,2}R^2V$ are dimensionless quantities, and $g_* = GM/R_*^2$ is the photospheric gravity (with stellar mass $M$ and radius $R_*$). Moreover,

$$\Xi = 4\dfrac{\Gamma(\alpha)}{1-\alpha}\,\Gamma_E\,\dfrac{\kappa_0 v_{th}}{\sigma_e c}, \tag{29}$$

where the photospheric Eddington factor $\Gamma_E = \sigma_e R_*^2 \pi I_*/cGM$ describes the ratio between the acceleration due to Thompson scattering on electrons (scattering coefficient $\sigma_e$) and $g_*$. Finally

$$\tilde{\beta}_L = \dfrac{1}{2} \int_{-1}^{1} d\mu \dfrac{1}{\tau_0} [(\tau_0 + \tau_0')^{1-\alpha} - \tau_0'^{1-\alpha}], \tag{30}$$

and similarly for $\tilde{\beta}_{Lc}$ (including an extra factor $D(\mu)$) and $\tilde{\gamma}_{Lc}$ (including an extra factor $\mu D(\mu)$). In the one-point coupling limit, $\tau_0' \to 0$, the above line force reduces to the well-known expression

$$g_L(r) = 4\pi \dfrac{\Gamma(\alpha)}{1-\alpha} \dfrac{\kappa_0 v_{th}}{c^2} \dfrac{1}{2} \int_{-1}^{1} d\mu\,\mu \bar{I}(r,\mu)\tau_0(r,\mu)^{-\alpha}. \tag{31}$$

Equation (28) holds for any $R$, $V$, and $\Pi$; appropriate values here are the stellar radius, escape speed from the star, and an estimate for the mean wind density. According to Eq. (28), the line force is specified by four dimensionless parameters, $\Gamma_E$, $\sigma_e cR\Pi/V$, $\alpha$, and $\kappa_0 v_{th}/\sigma_e c$. The maximum possible value of $g_L/g$ is about 1000, if all lines are optically thin (Abbott 1982). This corresponds to the oscillator value $Q = \Gamma(\alpha)^{\frac{1}{1-\alpha}} \kappa_0 v_{th}/\sigma_e c$ (Gayley 1995). In a dense wind, $g_L/g$ is typically between 1 and 10. The flow adjusts to a maximum mass loss rate, which implies small acceleration and strong self-shadowing of lines. The value of $\alpha$ lies between 1/2 and 2/3 for dense winds (Puls et al. 1996).

## 6. Numerical implementation

We have coded the above formalism in a rather dense Fortran program of about 1000 lines, featuring standard lambda iteration on a discrete, one-dimensional radial grid. We now turn to issues related to the numerical algorithm.

For each point $r \in [R_-, R_+]$, the two radial coupling locations $r_-(r)$ and $r_+(r)$ are required as integral bounds in Eq. (28). On a discrete grid, $v(j)$ and $v(k)$ at the resonances $r_-(j)$ and $r_+(k)$ of $r(i)$ differ slightly from $v(i)$. This can cause an absurd $|\mu| > 1$ in Eq. (12). In the neighborhood of kinks, even $|\mu| \gg 1$ is possible. A simple perturbation analysis finds a remedy. Let $r'$ and $v' = v$ be the exact radial resonance location and speed, and $r' \pm \epsilon$, $v' = v \pm \eta$ be grid neighbors; let $\epsilon > 0$. Inserting into Eq. (12), keeping only first-order terms in $\epsilon$ and $\eta$, and assuming $|r' - r| \gg \epsilon$, which is reasonable for nonlocal coupling, one finds

$$\mu^2 = 1 \pm \dfrac{2\eta r'^2}{(r'^2 - r^2)v}. \tag{32}$$

To assure $\mu^2 < 1$, it must hold that $\eta < 0$ ($\eta > 0$) if $r' > r$ ($r' < r$). Hence, from the two neighboring grid points of the exact resonance location, the one that obeys the *negative slope rule*, $(v - v')/(r - r') < 0$, has to be chosen.

Calculating the integral (17) by quadrature, we already find strong numerical oscillations of $S$ as a function of radius after the first iteration step. They are strongest near kinks, but extend over the whole three-point coupling domain. The corresponding oscillations in the line force can make the attached hydrodynamics code (see below) collapse. The oscillations are already found when evaluating the integral $\int dr'\,K(r,r')$ numerically, with kernel $K$ from Eq. (15). Their origin lies in the substitution $\mu \to r'$. Figure 3 shows, from left to right, the passage from $r < r_2$ (case II) to $r > r_2$ (case III). For case II, the radial coupling region between $r_2$ and $r_+$ has a nearly circular shape; it has a large opening angle but small radial extent. The number of radial mesh points (spheres) intersecting this resonance cap is small, and varies strongly with $r$, which leads to oscillations in the above integral and in $S(r)$. The opening angle of the resonance cap remains roughly constant as $r$ approaches $r_2$ (see Fig. 3). Therefore, the radiation energy emitted from the cap towards $r$ is roughly constant, even if $r$ is close to a kink, and the area of the cap is small. By contrast, the lateral coupling region $[r_1, r_2]$ of case II in Fig. 2 is resolved well by the $r'$ mesh. This part of the resonance surface is seen nearly edge-on from $r$, and thus contributes little to $S(r)$.

We conclude that a major drawback of the kernel formulation Eq. (28) for three-point coupling is that "dim" regions (lateral coupling) are oversampled by the radial mesh, whereas "bright" regions (radial coupling) are undersampled in the vicinity of kinks. This causes strong oscillations in $S$ over the whole radiative coupling domain. As a remedy, we combine Eqs. (11) and (17) to calculate $S_\nu(r)$ near kinks,

$$\beta(r)S_\nu(r) = \tilde{\beta}_c(r)I_\nu^* \pm \dfrac{\kappa v_{th}}{2} \int_{r_a}^{r_\pm} dr' r'^2 \varrho(r')S_\nu(r')K(r,r')$$
$$\pm \dfrac{1}{2} \int_{\mu_a}^{\pm 1} d\mu\,(1 - e^{-\tau'})\dfrac{1 - e^{-\tau}}{\tau}S_\nu(r'). \tag{33}$$

The $\mu$ integral covers the radial resonance cap extending from $\mu_a = \mu(r_a)$ to 1, and the $r'$ integral covers the lateral part of the resonance surface. For case II, the resonance cap extends to $r_a = r_1$ resp. $r_2$. For cases I and III without sharp turnover between the radial and lateral coupling region, we choose for $r_a$ some arbitrary point on the resonance surface near the kink. The $\mu$ integral is calculated either by semi-analytic approximation or by numerical quadrature. We find that this integral splitting gives better results than mesh refinement.

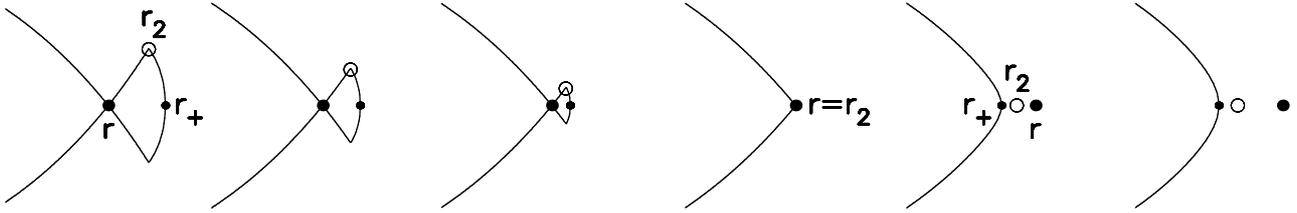

**Fig. 3.** Cut through the resonance surface for $r$ passing through the outer kink. In the left three panels (case II), $r_2$ separates the lateral from the radial coupling region.

There remain minor oscillations in $S$ and $g_L$, which we suppose are a consequence of $d^2v/dr^2 \to \infty$ at kinks. More importantly, $S(r)$ occasionally shows spikes at the kinks, which also seem to be a consequence of the discontinuity in $dv/dr$. The spikes lead to divergence of the $S$ iteration. To prevent this, we set $S = 0$ at two grid points on each side of both kinks and at the kinks themselves. This is justified since the one-point line force $g_L \sim dv/dr$ vanishes at kinks anyhow. An alternative method would be to smooth out the kinks.

## 7. Time-dependent wind model

We coupled our Fortran implementation of the RH algorithm to a time-dependent hydrodynamics code. The latter is a standard time-explicit, one-dimensional (assuming spherical symmetry) van Leer solver on a staggered mesh, using operator splitting for the continuity and the Euler equation (a similar method to the one used in the Zeus code by Stone & Norman 1992). The fundamental hydrodynamic quantities are the gas density $\varrho$ and radial momentum density $\varrho v$. The gas is assumed to be isothermal, $p = a^2\varrho$, with sound speed $a$. At the inner boundary, the photosphere, we fix $\varrho$ and extrapolate $\varrho v$ from the interior grid, corresponding to one incoming characteristic and one outgoing. At the outer boundary, both $\varrho$ and $\varrho v$ are extrapolated from the interior grid, corresponding to two outgoing characteristics.

As shown above, the wind model is fully specified by four dimensionless parameters. Since $\varrho$ and $dv/dr$ enter the line force nonlinearly (exponent $\alpha$), the wind adjusts to a stationary, monotonic eigenvalue solution. We adopt the following parameter values, appropriate for a dense wind from an O giant,

$$\Gamma_E = 0.3, \quad \sigma_e c \frac{R\Pi}{V} = 5, \quad \alpha = \frac{1}{2}, \quad \frac{\kappa_0 v_{\rm th}}{\sigma_e c} = 500. \quad (34)$$

The last two values correspond to $Q = 1570$.

We artificially enhance the stellar gravity $g \sim r^{-2}$ by a factor of 4 in the interval from $r/R_* = 1.5$ to $2$ at all simulation times. This introduces an effective wind nozzle that causes overloading (Abbott 1980), and allows us to study the general influence of multiple radiative coupling. A future paper will treat more realistic non monotonic flows, as caused by a sudden depression of the line force (due to ionization changes, density enhancements, etc.)

Equation (28) holds for three-point radiative coupling only, whereas the evolving wind velocity law may temporally show more than two kinks, hence more than three coupling locations. We enforce three-point coupling in the calculation of $S_e$ and $g_L$ by adopting only the innermost maximum and subsequent minimum of $v(r)$ as kinks. Any further non monotonicity of $v(r)$ is replaced in the RH code by a plateau with a small positive slope.

The numerical specifications of the wind model are as follows: 500 logarithmically spaced grid points are used from $r/R_* = 1$ to 5. Escape probabilities $\beta$ and $\beta_c$ are calculated using $100 \mu$ quadrature points. The source function is iterated until changes between subsequent iterations are below one tenth of a percent, which is the case after 3 to 4 iterations. In the vicinity of kinks (12 points each to the left and right), Eq. (33) is applied. Then $r_2$ for cases I and III is chosen to lie 10 grid points away from $r$. The one-point Sobolev approximation provides the initial values for $S$. A stationary, overloaded wind subject to the standard one-point Sobolev force is used as a starting model. A complete model run till $t = 10R_*/v_{\rm esc}$ (corresponding to more than two flow times over the mesh) requires more than 10,000 time steps (at a Courant number 0.1), or 1 cpu day, on a dedicated workstation, compared to a few cpu seconds for a one-point Sobolev model.

After passage through transient phases, the flow settles to a stationary solution. Figure 4 shows that the wind decelerates at all locations above the first kink where overloading sets in. This contrasts to the one-point model where the wind re-accelerates once overloading ceases.

Since the wind does not re-accelerate, case III does not exist. The resonance surface extends to infinity, and one has to apply an outer radiative boundary condition. To this end, we artificially steepen $v(r)$ on the last mesh points, to reach a speed slightly above that at the inner kink. This closes the resonance surface within the calculational domain. The source function is put to zero in this region, suppressing any radiation entering from larger radii.

Figure 4 shows that the terminal wind speed is roughly half the escape speed. The outer wind density is larger by a factor of two than in the start model. The line force in the outer wind drops by a factor of six compared to the start model, to almost balance with gravity.

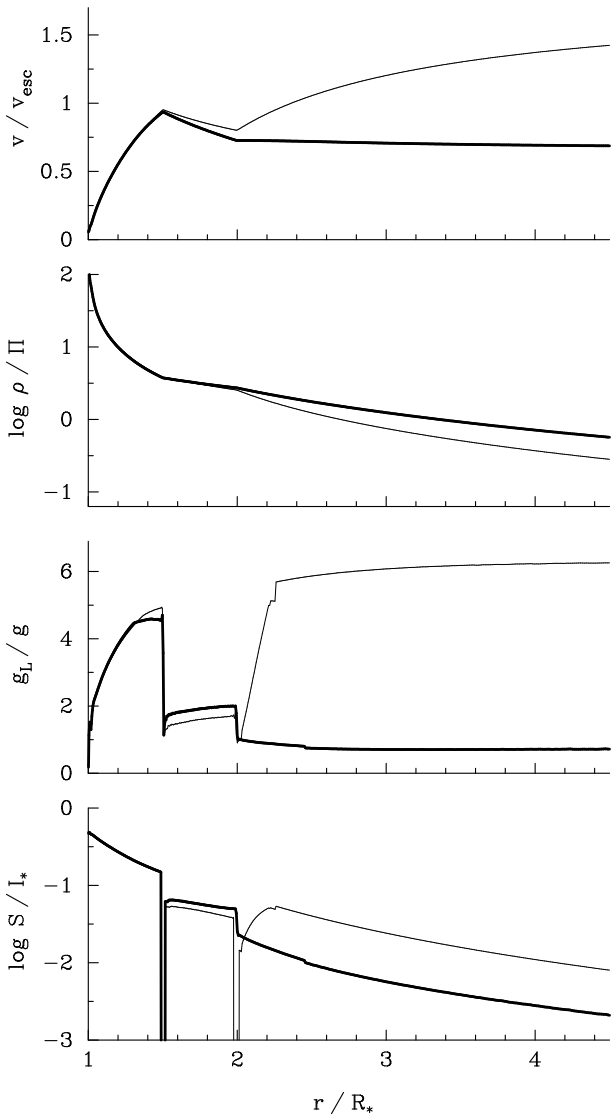

**Fig. 4.** From top: velocity, density, line force, and source function in an overloaded wind subject to three-point radiative coupling, as a function of radius. Only the initial one-point Sobolev start model (thin line) and the converged, stationary solution (heavy line) at dimensionless time $tv_{\rm esc}/R_* = 10$ are shown.

## 8. Comparison with other methods

The RH approach is complementary to the smooth source function method (SSF: Owocki 1991; Owocki & Puls 1996) used in time-dependent hydrodynamic simulations of unstable hot star winds (Owocki et al. 1988; Feldmeier 1995). The instability originates in line deshadowing (Lucy & Solomon 1970; Carlberg 1980; Owocki & Rybicki 1984; Lucy 1984), which makes it necessary to resolve the line profile in the calculation of the radiative line force. The Sobolev approximation is not appropriate here. In the diffuse flux

$$\bar{H}_{\nu,{\rm d}}(r) = \frac{1}{2}\int_{-1}^{1} d\mu\,\mu \int_{-\infty}^{\infty} dx\,\phi(x) \int d\tau'\,S(r')e^{-|\tau(x,\mu,r)-\tau'|} \quad (35)$$

entering the line force, one assumes $S(r') \approx S(r)$ and takes $S$ out of the integral. The remaining $\tau'$ integral is readily solved. For $S(r)$, simple, local estimates from Sobolev approximation are used. The optical depth $\tau$ is a fast-varying function at resonance locations, hence the accurate calculation of $\tau$ on the highly resolved $r$ and $x$ meshes is the major cpu time-consumer in instability simulations. Since $S$ is a local quantity, only $\tau$ carries information about nonlocal radiative coupling (shadowing).

The RH approach adopted in the present paper follows the opposite strategy, namely to use simple estimates for $\tau$ from Sobolev approximation, and to iterate the source function $S$ in order to account for nonlocal radiative coupling. Future instability calculations will therefore largely benefit from merging the two methods, thus accounting for nonlocal coupling both in $S$ and $\tau$.

Baron & Hauschildt (2004) developed a short characteristics method for solving the spherical symmetric radiative transfer equation for non monotonic flows using accelerated lambda iteration in the comoving frame. Their approach successfully accounts for the full boundary value problem both in the wavelength and spatial dimensions, and these authors plan to incorporate the technique in a full three-dimensional radiation hydrodynamics code. So far it is applied to fixed, prespecified one-dimensional velocity laws, without coupling to a hydrodynamics solver. From Table 1 in Baron & Hauschildt, their method seems to be roughly a factor of 30 slower than ours, which should be largely due to intricate matrix calculations.

For the future it would be interesting to directly compare results from the ALI method of Baron & Hauschildt (2004) with the present Sobolev approach. The latter is accurate for flows with steep velocity gradients, where the Sobolev length is shorter than the hydrodynamics lengthscale, and can be applied to the examples given in Baron & Hauschildt.

## 9. Summary

In the present paper we extend the geometry-free formalism of Rybicki and Hummer (1978) that allows to calculate the source function and radiative line force in the presence of three-point radiative coupling in spherically symmetric winds. The strong influence of three-point coupling on the dynamics of radiation-driven winds with non monotonic velocity laws is demonstrated in a simplified model of an overloaded (choked) flow. For the first time to our knowledge, multiple radiative coupling is accounted for in a time-dependent hydrodynamic wind simulation. The resulting wind decelerates at all radii above the point where overloading sets in. This contrasts to standard models with a purely local radiative force, where the wind starts to accelerate again.

The coasting wind solution in the multiple coupling case results from a sharp drop-off in the source function $S$ and the line force $g_{\rm L}$ at the resonance surface. This drop-off should be a robust feature and not depend on specific assumptions of the present simulation. We conclude that radiation-driven winds that start to decelerate at a certain radius (due to choking, ion decoupling, instability, etc.) decelerate at *all* larger radii. According to Fig. 4, the terminal wind speed could be significantly reduced compared to standard, one-point coupling models.

We expect, however, that the deshadowing instability counteracts such a reduction in terminal speeds. The instability causes clumping, non monotonic velocity laws, and multiple resonances in the predominantly radial wind. The clumps (and thus the resonance surfaces) will have a small lateral extent and random radial spacing due to the small coherence length and time of turbulent seed perturbations, respectively. In the deepest shadow behind the radial cap of the resonance surface, $g_L$ drops off and radiation driving ceases. However, photons propagating non-radially penetrate into the shadowed region and re-accelerate the wind. This has to be clarified in future two-dimensional simulations.

*Acknowledgements.* We appreciate discussions with Wolf-Rainer Hamann, Leon Lucy, Stan Owocki, and Joachim Puls. This work was supported by the DFG under grant number Fe 573/2-1.